\documentclass[11pt]{article}
\usepackage{hyperref}
\usepackage{amssymb}
\usepackage{graphicx}
\usepackage{slashed}
\usepackage{amsmath}
\usepackage{breqn}
\usepackage{authblk}
\usepackage{float}
\usepackage{dsfont}
\usepackage{cite}
\usepackage[section]{placeins}
\usepackage[a4paper, total={6.8in, 10in}]{geometry}
\numberwithin{equation}{section}
\newcommand{\be}{\begin{equation}}
\newcommand{\ee}{\end{equation}}
\newcommand{\bea}{\begin{eqnarray}}
\newcommand{\eea}{\end{eqnarray}}
\newcommand{\ba}{\begin{aligned}}
\newcommand{\ea}{\end{aligned}}
\begin{document}
\title{Casimir effect in DFR space-time}

\author{E. Harikumar \thanks{eharikumar@uohyd.ac.in} and Suman Kumar Panja \thanks{sumanpanja19@gmail.com}}
\affil{School of Physics, University of Hyderabad, \\Central University P.O, Hyderabad-500046, Telangana, India}

\maketitle

\begin{abstract}
Non-Commutative space-time introduces a fundamental length scale suggested by approaches to quantum gravity. Here we report the analysis of the Casimir effect for parallel plates separated by a distance of $L$ using a Lorentz invariant scalar theory in a non-commutative space-time (DFR space-time), both at zero and finite temperatures. This is done in two ways; one when the additional space-dimensions introduced in DFR space-time are treated as extra dimensions but on par with usual space-dimension and in the second way, the additional dimensions are treated as compact dimensions. Casimir force obtained in the first approach coincides with the result in the extra-dimensional commutative space-time and this is varying as $\frac{1}{L^5}$. In the second approach, we derive the  corrections to the Casimir force, which is dependent on the separation between the plate, $L$ and on the size of the extra compactified dimension, $R$. Since correction terms are very small, keeping only the most significant terms of these corrections, we show that for certain values of the R, the corrections due to non-commutativity  makes the force between the parallel plates more attractive, and using this, we find lower bound on the value of $R$. We show here that the requirement of the Casimir force and the energy to be real, impose the condition that the weight function used in defining the DFR action has to be a constant. At zero temperature, we find correction terms due to non-commutativity, depend on $L$ and $R$ dependent modified Bessel functions $K_{1}$ and  $K_{2}$, with coefficients that vary as $\frac{1}{LR^3}$ and $\frac{1}{L^2R^2}$, respectively . For finite temperature, the Casimir force has correction terms that scale as $\frac{1}{L}$ and $\frac{1}{L^3}$ in high-temperature limit and as $\frac{1}{L^2}$ and $\frac{1}{L^4}$ in the low-temperature limit.

\end{abstract}

\section{Introduction}
Though it is the oldest known force, the nature of gravity is still unknown on microscopic scale. Understanding the nature of gravity at the quantum regime is one of the pressing issues in Physics. Many approaches have been developed to study quantum theory of gravity. Some of these approaches are string theory, loop gravity, emergent gravity, and non-commutative geometry\cite{connes,douglas,Glikman,dop,am,madore,seiberg-witten}. All of these approaches introduce a fundamental length scale, below which quantum effect of gravity comes into play\cite{Glikman,dop}. Non-commutative geometry provides a path to incorporate such a fundamental length scale\cite{connes,douglas,Glikman,dop}. Non-commutative geometry also appears in the low energy limit of string theory \cite{douglas}. In low energy limit, quantum gravity model based on spin-form reduce to a field theory in a non-commutative space-time known as $\kappa$-space-time \cite{livin}. 

In recent times non-commutative space-times and models on such space-times are being investigated intensively. One of the initial motivations behind introducing non-commutative space-time was to remove divergence from QFT. In \cite{snyder}, non-commutative space-time was proposed, but it was shown that the divergences were not completely removed in this non-commutative theory\cite{yang}. Research activities in different areas, such as non-commutative geometry, string theory, fuzzy sphere, etc., rekindled the interest in non-commutative space-times and study of physics on such space-times\cite{connes,douglas,Glikman,dop,am,madore,seiberg-witten,jabb}. One of the well studied non-commutative space-time is the Moyal space-time, whose coordinates satisfy
\be
[\hat{x}^{\mu},\hat{x}^{\nu}]=i\theta^{\mu\nu}, \nonumber \label{intro1}
\ee
where $\theta^{\mu\nu}$ is antisymmetric, constant tensor with dimension of $(length)^2$. This constant $\theta^{\mu\nu}$ leads to the violation of Lorentz invariance in field theory \cite{douglas} and breaking of rotational symmetry in non-relativistic theory. The violation of Lorentz symmetry makes definition of particles problematic and also it brings vacuum birefringence in theory\cite{khoz}.

To recover Lorentz symmetry in non-commutative theory, a transformation rule has been assigned to 
the non-commutative parameter $\theta^{\mu\nu}$ under Lorentz transformation\cite{dop} and further the components of the non-commutative parameter $\theta_{\mu\nu}$ have been elevated to coordinates of the space-time, thereby increasing the dimension of the underlying space-time\cite{carlson}. This space-time is called DFR space-time \cite{dop} and associated symmetry algebra is known as DFR algebra. The coordinates of the phase space corresponding to DFR space-time are $x_{i}$, $p_{i}$, $\theta_{i}$ and $k_{i}$. To get complete Fourier decomposition of field and to maintain Lorentz invariance in DFR space-time, canonical conjugate momentum corresponding to non-commutative coordinate $\theta^{\mu\nu}$ have been introduced \cite{amo1,amo2,amo3,amo5}. The associated symmetry algebra is called DFRA algebra. 

Since quantum theories of non-commutative space-time have an inherent feature of introducing length parameters (along with nonlocality and nonlinearity), it is of intrinsic interest to study the implications of this length scale in physical phenomena. One such phenomenon where length scale plays a significant role in the commutative space-time is the Casimir effect\cite{hbg, Mil}. It was shown that when 
two conducting plates are placed parallel to each other at a small separation, they experience an attractive force, and this force depends on the separation between the plate\cite{Mil}. It was shown that the zero-point energy of the electromagnetic field changes when the plates are introduced and the attractive force arises due to the change in the zero-point energy. The boundary condition imposed on the electromagnetic field by the introduction of plates 
is different from that in their absence, and this changes the zero-point energy in two situations. This change in the zero-point energy was shown to result in an attractive force between parallel conductors in \cite{hbg}.
Subsequently, this effect was studied by calculating the change in the zero-point energy of various quantized fields such as a real scalar field and fermionic field, for conductors with different geometries \cite{kam,Mil}. 
 The electromagnetic field at either side of the plates consists of two modes (TE and TM), while the real scalar field has only one degree of freedom but obeys the same boundary conditions as the electromagnetic field. 
 The Casimir energy and force calculated using the scalar field was shown to be half of the corresponding values obtained by considering the electromagnetic field, and this numerical factor difference is due to the differences in the number of modes associated with the scalar and the electromagnetic fields\cite{Mil,kam,pule}. 
 
  Modifications to the Casimir effect due to different dimensions of space-time, due to the presence of medium between plates, due to thermal effect, etc., using real scalar field theory have been studied\cite{kam,Brevik}. Experimental investigations of Casimir force for separation of few micrometers between the plates have been reported \cite{onofr,brax,sedmik}. Apart from the experimental study for parallel conducting plates, the Casimir effect has been investigated for other geometrical configurations \cite{sed,Bimonte,Wang}. Using these experimental results, constraints on the Yukawa type correction to Newtonian gravity are obtained \cite{onfi}.

In \cite{casadio,fosco,jab,pinto,skp}, Casimir effect has been studied in different non-commutative space-times such as Moyal space-time \cite{douglas} and $\kappa$-deformed space-time \cite{luk}, using scalar field theory. For even-dimensional Moyal space-time, using coherent state approach and smeared boundary condition, Casimir force is calculated between two parallel conducting plates \cite{casadio,fosco} using scalar field theory. In \cite{jab}, Casimir energy in 2+1 dimensional non-commutative space-time with non-trivial topologies was derived and analyzed. In \cite{gom,nam}, the Casimir effect on compact non-commutative space was studied. In \cite{pinto,skp}, the Casimir effect was investigated for $\kappa$-deformed scalar theory. Casimir effect was investigated using the Green's function approach, and a bound on the non-commutative length scale was calculated by comparing it with experimental results\cite{skp}. In $\kappa$-space-time, space coordinates commute among themselves, but commutation relation between time coordinate and space coordinates vary with space coordinates, i.e., they satisfy 
\be
[x^i, x^j]=0,~~~[x^0, x^i]=ax^i,~~~a=\frac{1}{\kappa}.
\ee
where $a$ is the deformation parameter. This indicates that as  in Moyal space-time, in the $\kappa$ space-time also the Lorentz symmetry is broken. But this problem is avoided in DFR space-time with introduction of transformation for the non-commutative parameter $\theta^{\mu\nu}$ \cite{dop}.
Latter, components of $\theta_{\mu\nu}$ have been promoted to coordinates, enlarging the dimension of the space-time\cite{carlson}. In commutative space-time, Casimir force in presence of extra dimensions has been explored in recent times\cite{kam,prd79,plb668,farina,teo}. Thus it is of intrinsic interest to study the Casimir effect in DFR space-time which has extra dimensions.

In the commutative space-time, Casimir effect between two parallel plates has been evaluated by modelling the plates with two $\delta$-function potentials at $x=0$ and $x=L$ \cite{bordag,Mil}, respectively. Casimir force on the plate at $x=L$ has been derived by taking difference of vacuum expectation values of Energy-Momentum tensor on either side of the plate. For massless scalar field theory, $xx$ component of Energy-Momentum tensor was found at a point, just left to the plate at $x=L$, i.e., $\hat{T}^{xx}\Big|_{x=L^{-}}$ and at another point, just right to the plate at $x=L$, i.e., $\hat{T}^{xx}\Big|_{x=L^{+}}$. Then force was calculated at $x=L$ by taking difference between the vacuum expectation value of the stress-tensor, in the strong interaction limit of $\delta$-function potential as,
\be
\hat{F}_{\gamma,\gamma'\rightarrow \infty}=<\hat{T^{xx}}>\Big|_{x=L^{-}}-<\hat{T^{xx}}>\Big|_{x=L^{+}}. \label{one} 
\ee
One way to derive vacuum expectation value of stress tensor was by rewriting Energy-Momentum tensor as an operator acting on the vacuum expectation value of the time ordered product of fields at nearby points, $x$ and $x'$ as,
\bea
<\hat{T}^{\mu\nu}_{x,x'}> &=& <\hat{O}^{\mu\nu}(\partial,\partial')T(\phi\phi')> \nonumber \\
&=& \hat{O}^{\mu\nu}(\partial,\partial')<T(\phi \phi')> \nonumber \label{two}
\eea
where $\hat{O}^{\mu\nu}(\partial,\partial')$ is a derivative operator, $\phi=\phi(x)$ and $\phi'=\phi(x')$. Since 
\be
<T(\phi(x)\phi(x'))>=iG(x,x'), \label{three}
\ee
where $G(x,x')$ is the Green's function. One can relate the vacuum expectation value of the Energy-Momentum tensor to the Green's function. Thus to calculate RHS of Eqn.(\ref{one}), Green's functions in different regions are first derived by solving Euler-Lagrange equation of scalar field theory with Dirichlet boundary condition and using these Green's functions, Casimir force is obtained. Casimir energy is calculated by integrating the Casimir force over the separation distance between two parallel plates, as
\be
E_{\gamma,\gamma'\rightarrow \infty}=-\int \hat{F}_{\gamma,\gamma'\rightarrow \infty}dx \label{four}
\ee
where $\gamma$ and $\gamma'$ are coupling strengths of $\delta$-function potentials, modelling the plates and $\gamma,\gamma' \rightarrow \infty$ denote the strong interaction limit. The Casimir force and energy calculated using the scalar field differs from the corresponding results obtained using electromagnetic field by a factor of $\frac{1}{2}$.

In this paper, we study the Casimir effect between two parallel plates in DFR space-time by analyzing DFRA scalar field theory \cite{amo1,amo2,amo3,amo5}. Since the non-commutative coordinates do not show up in the commutative limit, and our aim is to investigate the corrections to the Casimir force in the commutative space-time, we consider the situation where the plates are kept only in the x-directions (commutative directions). We start with the action of $10$-dimensional DFRA scalar field theory. For unitarity of this field theory one sets temporal part of non-commutative coordinate, $\theta^{\mu\nu}$ to be vanishing, i.e., $\theta^{0i}=0$ and further we take definitions, $\theta^{i}=\frac{1}{2}\epsilon^{ijl}\theta_{jl}$ and $k^{i}=\frac{1}{2}\epsilon^{ijl}k_{jl}$ \cite{abreu1,gomis,chaichian}. After these, we get the action of $7$-dimensional DFRA scalar field theory. We start our study from the Lagrangian of DFRA scalar field theory with plates modelled through $\delta$-function potentials, as in the commutative space-time. Next, we vary the corresponding action and derive the general form of the Euler-Lagrange equation and components of the Energy-Momentum tensor. Without loss of generality, we restrict our attention to $4+1$ dimensions $(x^{\mu},\theta^{1})$, and study the Casimir effect in two ways. In first way, we treat extra $\theta$-dimension due to noncommutativity of the $4+1$-dimensional DFR space time, in the same footing as transverse directions, $y$ and $z$. Then following the standard calculational procedure given in \cite{kam,Mil} we show that the Casimir force  vary as $\frac{1}{L^5}$, where $L$ is the distance between the parallel plates. This force expression exactly coincides  with result given in \cite{Mil} for extra-dimensional commutative space-time. In second approach, we compactify the new extra dimension $\theta$ introduced by noncommutativity in DFR space-time. After compactifying the extra dimension, we solve the equation of motion using Dirichlet boundary condition and get Green's functions in different regions. We also derive the $xx$ component of the Energy-Momentum tensor at either side of the plate at $x=L$. Then we obtain the Casimir force by taking the difference between vacuum expectation values of stress tensors at $x=L$ as shown in Eqn.(\ref{one}) for the commutative case. For this, we use the relation between Green's function and time-ordered product of fields in the nearby points in $4+1$-dimensional DFR space-time, viz;
\be
<T(\phi(x,\eta)\phi(x',\eta'))>=iG(x,x';\eta,\eta'), \label{five}
\ee
where $G(x,x';\eta,\eta')$ is the Green's function obtained from Euler-Lagrange equations. Here $\eta=\frac{\theta}{R}$ is compactified direction and R is the size of the extra compactified dimension. Using this, we calculate Casimir force and derive Casimir energy. Further, we study finite temperature corrections to the Casimir effect for DFRA real scalar field theory and obtain Casimir force in the high temperature and low-temperature limits. We also show that our results reduce to commutative space-time results in the appropriate limit of the size of the extra compactified dimension. We show that the Casimir force gets two types of corrections due to non-commutativity of space-time, one is $\frac{L}{R}$ dependent modified Bessel function-$K_{1}$ whoes coefficient vary as $\frac{1}{L}$ and the other is $\frac{L}{R}$ dependent modified Bessel function-$K_{2}$ whose coefficient vary  as $\frac{1}{L^2}$. In the high-temperature limit, we find the correction terms that vary as $\frac{1}{L}$ and $\frac{1}{L^3}$ while in the low-temperature limit Casimir force has correction terms that vary as $\frac{1}{L^2}$ and $\frac{1}{L^4}$. We also show that the modifications to the Casimir force in both zero temperature and the finite temperature, has dependency on $R$, size of the extra compactified dimension. To understand the generic effect of the non-commutativity on the Casimir force, we then consider most significant correction terms and analyzing the plots for the Casimir force with these correction terms, we obtain constraints on the size of the extra compactified dimension $R$. 

We also show that the massless DFRA scalar theory in 7-dimensions gives a massive scalar theory in 4-dimensions under Kaluza-Klein reduction. Also the modification of the Casimir effect using this model is calculated.

This paper is organized as follows. In the next section, we briefly discuss DFR space-time and underlying DFRA algebra \cite{amo1,amo2,amo3,amo5}. In Sec.3., we derive the general form of the Euler-Lagrange equation, current density, and the components of Energy-Momentum tensor corresponding to the scalar theory on the $7$-dimensional DFR space-time. Our main results are obtained in Sec.4. Here we construct the Lagrangian describing massless DFRA scalar field theory in the presence of two parallel plates. We impose the Dirichlet boundary condition and calculate the Casimir force and the energy for massless scalar field with parallel plates in $4+1$-dimensional DFR space-time. The $\theta^{\mu\nu}$ dependent terms in the Lagrangian captures the effects of non-commutativity of space-time. First, we study the Casimir effect by treating the new extra dimension-$\theta$ introduced by non-commutativity in $4+1$ dimensional DFR space-time, in same footing with other two transverse commutative space dimensions- $y$ and $z$. Then by following the standard procedure, we show that the Casimir force expression is exactly same as the result obtained in case of commutative extra dimensional space-time. Then in Subsection.4.1, we study the Casimir effect by treating the additional dimension-$\theta$ as a compactified dimension in $4+1$ dimensional DFR space-time. This is carried out by first compactifying the extra dimension $\theta$ of DFR space-time. Then we derive Green's function for different regions of interest by solving the Euler-Lagrangian equation, describing the interaction of scalar field with parallel plates in presence of extra compactified dimension in $4+1$-dimensional DFR space-time. Next we obtain the energy-momentum tensor corresponding to the massless DFRA real scalar theory from the general form of the stress-tensor constructed in Sec.3. Then we derive the vacuum expectation value of the energy-momentum tensor in terms of the obtained Green's function. Using the reduced Green's function obtained in different regions, we calculate the Casimir force and Casimir energy in presence of extra compactified dimension-$\theta$ for $4+1$ dimensional, massless DFRA scalar field theory. Further, we investigate finite temperature modifications to the Casimir effect in this model in Sec.5. We also study corrections to Casimir force for both the low-temperature limit and the high-temperature limit. Our concluding remarks are given in Sec.6. In the appendix, we discuss the study of the Casimir effect for massive scalar theory in $4$-dimensional Minkowski space-time, obtained using Kaluza-Klein dimensional reduction from a $7$-dimensional, massless, DFRA complex scalar theory. We use the metric, $\eta_{AB}=(\eta_{\mu\nu},-1)=diag(+1,-1,-1,-1,-1)$. 

\section{DFR space-time}
In this section, we present a brief summary of the DFR space-time, a non-commutative space-time that respects Lorentz invariance and also the symmetry algebra associated to this space-time\cite{dop}. The Moyal-space-time coordinates satisfy
\be 
[\hat{x}_{\mu},\hat{x}_{\nu}]=i{\theta}_{\mu\nu}. \label{moyal}
\ee
Since $\theta_{\mu\nu}$ is a constant antisymmetric tensor, it results in the violation of Lorentz symmetry \cite{douglas}. Violation of Lorentz invariance brings issues with the interpretation of particle-antiparticle states. Assigning an appropriate transformation to $\theta_{\mu\nu}$ under Lorentz transformation reinstates the Lorentz invariance of theory \cite{dop} and thus avoids the issue of defining particle-antiparticle states. In \cite{carlson}, the components of the $\theta_{\mu\nu}$ were promoted to coordinate operators $\hat{\theta}_{\mu\nu}$. Thus the non-commutative space-time gets six more dimensions and the resulting 10-dimensional space-time is known as DFR space-time. The corresponding algebra is given by
\be
\begin{split}
[\hat{x}_{\mu},\hat{x}_{\nu}]&=i\hat{\theta}_{\mu\nu},~~[\hat{x}_{\mu},\hat{\theta}_{\nu\rho}]=0,\\
[\hat{x}_{\mu},\hat{p}_{\nu}]&=i\eta_{\mu\nu},~~[\hat{\theta}_{\mu\nu},\hat{\theta}_{\rho\lambda}]=0 \\
[\hat{p}_{\mu},\hat{\theta}_{\nu\lambda}]&=0,~~~~~~[\hat{p}_{\mu},\hat{p}_{\nu}]=0 \label{algebra1}
\end{split}
\ee
For Fourier decomposition of field, and also for the Lorentz invariance of the field theory defined on this non-commutative space-time, one needs to introduce conjugate momentum $\hat{k}_{\mu\nu}$ - corresponding to $\hat{\theta}_{\mu\nu}$, apart from $\hat{x}^{\mu}$, $\hat{p}_{\mu}$ and $\hat{\theta}_{\mu\nu}$. In DFR space-time \cite{amo1,amo2,amo3}, non-commutative coordinates $\hat{\theta}_{\mu\nu}$ and their canonical conjugate momenta $\hat{k}_{\mu\nu}$ are considered as operators alongside with the commutative coordinates, $\hat{x}_{\mu}$ and their canonical conjugate momenta $\hat{p}_{\mu}$ . Thus algebra is enlarged by
\be
\begin{split}
[\hat{x}_{\mu},\hat{k}_{\nu\lambda}]&=-\frac{i}{2}(\eta_{\mu\nu}\eta_{\rho\lambda}-\eta_{\mu\lambda}\eta_{\nu\rho})\hat{p}^{\rho},~~[\hat{p}_{\mu},\hat{k}_{\nu\lambda}]=0,\\
[\hat{\theta}_{\mu\nu},\hat{k}_{\rho\lambda}]&=i(\eta_{\mu\rho}\eta_{\nu\lambda}-\eta_{\mu\lambda}\eta_{\nu\rho}),~~
[\hat{k}_{\mu\nu},\hat{k}_{\rho\lambda}]=0, \label{algebra2}
\end{split}
\ee
in addition to the ones given in Eqn.(\ref{algebra1}).
The relations in Eqn.(\ref{algebra1}) and Eqn.(\ref{algebra2}) together is a closed algebra-DFRA algebra\cite{amo1,amo2,amo3,amo5}.
 The corresponding Lorentz generator is defined as \cite{amo1,amo5}
\be 
M_{\mu\nu}=\hat{X}_{\mu}\hat{p}_{\nu}-\hat{X}_{\nu}\hat{p}_{\mu}-\hat{\theta}_{\mu\lambda}{\hat{k}}_{\nu}~^{{\lambda}}+\hat{\theta}_{\nu\lambda}{\hat{k}}_{\mu}~^{\lambda} \label{dlorentz}
\ee
where,
\be 
\hat{X}_{\mu}=\hat{x}_{\mu}+\frac{1}{2}\hat{\theta}_{\mu\nu}\hat{p}^{\nu}. \label{X}
\ee 
The shifted coordinates in the above equation satisfy commutation relations
\be 
[\hat{X}_{\mu},\hat{X}_{\nu}]=0,~~~[\hat{X}_{\mu},\hat{p}_{\nu}]=i\eta_{\mu\nu}. \label{X1}
\ee
With the relations given in Eqn.(\ref{algebra1}), Eqn.(\ref{algebra2}) and Eqn.(\ref{dlorentz}), one sees that the generators of DFRA algebra are closed, i.e.,
\be 
\begin{split}
[M_{\mu\nu},\hat{p}_{\lambda}]&=i(\eta_{\mu\lambda}\hat{p}_{\nu}-\eta_{\nu\lambda}\hat{p}_{\mu}),\\
[M_{\mu\nu},\hat{k}_{\alpha\beta}]&=i(\eta_{\mu\beta}\hat{k}_{\alpha\nu}-\eta_{\mu\alpha}\hat{k}_{\nu\beta}+\eta_{\nu\alpha}\hat{k}_{\beta\mu}-\eta_{\nu\beta}\hat{k}_{\alpha\mu}),\\
[M_{\mu\nu},M_{\lambda\rho}]&=i(\eta_{\mu\rho}M_{\nu\lambda}-\eta_{\nu\rho}M_{\lambda\mu}-\eta_{\mu\lambda}M_{\rho\nu}+\eta_{\nu\lambda}M_{\rho\mu}), \label{dlorentz1}
\end{split}
\ee
The above relations of DFRA algebra are consistent with the Jacobi identities (see \cite{amo1,amo5}). The Casimir operator associated with the DFRA algebra is given as \cite{amo1}
\be 
\hat{P}^2=\hat{p}_{\mu}\hat{p}^{\mu}+\frac{\lambda^2}{2}\hat{k}_{\mu\nu}\hat{k}^{\mu\nu}, \label{casimir}
\ee
where $\lambda$ is the non-commutative parameter with the length dimension. In \cite{amo1,amo2}, the construction of infinitesimal transformations of $x_{\mu},~\theta_{\mu\nu},~p_{\mu},~k_{\mu\nu}$ and $M_{\mu\nu}$ and their relevance for the DFR space-time are discussed in detail.



\section{ Construction of general form of Energy-Momentum tensor and Current density}
One way to evaluate the Casimir force experienced by a plate is to obtain the vacuum expectation value of the Energy-Momentum tensor on either side of this plate and find their difference. Since the Energy-Momentum tensor can be written as a differential operator acting on the product of fields, this vacuum expectation value of the Energy-Momentum tensor is related to the vacuum expectation value of the time-order product of fields at two space-time points acted upon by this operator. This allows one to express vacuum expectation value of Energy-Momentum tensor as the operator acting on Green's function, i.e.,
\bea
<T_{x,x'}^{\mu\nu}> &=& \hat{O}<T(\phi(x)\phi(x'))> \nonumber \\
&=& \hat{O}G(x,x') \label{ten-green}
\eea 
The Green's function on either side of the plate are calculated by solving the Euler-Lagrange equation for these different regions. 
In this section, we summarize the construction of action for the scalar field theory in DFR space-time using the definition of the Moyal star product \cite{douglas}. By varying the action, we then obtain the equations of motion corresponding to the scalar field defined in the DFR space-time. Here we also discuss the implications of the $\theta$-dependent weight function introduced in the definition of action. 
\subsection{Construction of Action}
Before construction of the action for the DFRA scalar field, we start with the definition of star product \cite{dop}, which for two functions $f$ and $g$ of $x_{\mu}$ and $\theta_{\mu\nu}$, given as 
\be
f(x,\theta)\star g(x,\theta)=e^{\frac{i}{2}\theta^{\mu\nu}\partial_{\mu}\partial_{\nu}'}f(x,\theta)g(x',\theta)\Big|_{x=x'}. \label{starproduct}
\ee 
Using above definition, it can be shown that the Moyal product satisfy following relation,
\be
\int d^4x~d^6\theta~W(\theta)~f(x,\theta)\star g(x,\theta)=\int_{-\infty}^{\infty} d^4x~d^6\theta~W(\theta)~f(x,\theta)g(x,\theta). \label{spproduct}
\ee
$W(\theta)$ in the above equation is the weight function and has been introduced to avoid divergence in perturbative QFT on DFR space-time \cite{kase,carlson,imai,saxell}. For Lorentz invariance, $W(\theta)$ is taken as even function of $\theta$ and properties of the weight function is discussed in \cite{kase,carlson,imai,saxell} in detail. Here we consider weight function to be Gaussian function, given by
\be
W(\theta)=\Big(\frac{1}{4\pi^2\lambda^4}\Big)^{3}e^{-\frac{\theta^2}{8\lambda^4}}, \label{weight}
\ee
where $\lambda$ is the non-commutative parameter of length dimension. 

Now one sets up the action for the scalar field theory in DFR space-time by replacing the usual product in the Lagrangian with the star product. As a result, the non-commutative action for the DFRA scalar field is
\be
S=\int d^4x~d^6\theta~W(\theta)\frac{1}{2}\Big(\partial_{\mu}\phi\star\partial^{\mu}\phi+\frac{\lambda^2}{2}\partial_{\theta^{\mu\nu}}\phi\star\partial^{\theta^{\mu\nu}}\phi-m^2\phi\star\phi\Big). \label{action1}
\ee
After using the property given in Eqn.(\ref{spproduct}), the action for DFRA scalar field becomes
\bea
S = \int d^4x~d^6\theta~W(\theta)\frac{1}{2}\Big(\partial_{\mu}\phi\partial^{\mu}\phi+\frac{\lambda^2}{2}\partial_{\theta^{\mu\nu}}\phi\partial^{\theta^{\mu\nu}}\phi-m^2\phi^2\Big). \label{action2}
\eea 
For constant value of weight function (specifically for $W(\theta)=1$) above action reduces to that studied in \cite{amo5}.

In order to preserve the unitarity of the field theory on DFR space-time, it has been shown that one needs to set temporal part of non-commutative coordinates to be vanishing, i.e., $\hat{\theta}_{0i}=0$ \cite{gomis} and thus the resulting DFR space-time has seven dimensions. Further we use definitions, $\theta^{i}=\frac{1}{2}\epsilon^{ijl}\theta_{jl}$ and $k^{i}=\frac{1}{2}\epsilon^{ijl}k_{jl}$ \cite{abreu1} and thus from above we find $4+3$ dimensional action for real scalar field in DFR space-time as
\bea
S &=& \int d^4x~d^3\theta~W(\theta){\cal L}(x,\theta,\phi,\partial_{\mu}\phi ,\partial_{\theta^{i}}\phi) \label{gen-action1} \\ 
& = &\int d^4x~d^3\theta~W(\theta)\frac{1}{2}\Big(\partial_{\mu}\phi\partial^{\mu}\phi+\lambda^2 \partial_{\theta^{i}}\phi\partial^{\theta^{i}}\phi-m^2\phi^2\Big). \label{action3}
\eea
Note here that the $W(\theta)$ in the measure and $\lambda$ dependent term in Lagrangian bring in the non-commutative contribution due to the DFR space-time. We get the commutative action of the scalar field through compactification i.e., for $\lim_{\lambda\to 0}W(\theta)=\lim_{\lambda\to 0}\frac{e^{-\frac{\theta^2}{4\lambda^4}}}{2\pi\lambda^2}=\delta(\theta)$ and $\int d\theta~W(\theta)=1$. Note that in \cite{amo5}, $W(\theta)$ is taken to be equal to 1.
\subsection{Construction of Energy-Momentum tensor and Current density for DFRA scalar field}
Here in this subsection, we obtain the expression of Energy-Momentum tensor and the current density for DFRA scalar field theory by varying the action by an infinitesimal change in the space-time coordinates of DFR space-time, $(x^{\mu},\theta^{\mu\nu})$.

We start with the general form of action (\ref{gen-action1}),
\be
S = \int_{R} d^4x~d^3\theta~W(\theta){\cal L}(x,\theta,\phi,\partial_{\mu}\phi ,\partial_{\theta^{i}}\phi)
\ee
where R is a bounded region.
Next we vary the action by considering functional change of field as well as infinitesimal change in space-time co-ordinates such as ${x^{\mu}}^{\prime}=x^{\mu}+\epsilon\delta x^{\mu}$ and ${\theta^{i}}^{\prime}=\theta^{i}+\epsilon \delta{\theta^{i}}$ and thus we get
\bea
\delta S &=&\int_{R} d^4x'd^3\theta' W(\theta')\mathcal{L'}(\phi',\partial_{\mu}\phi',\partial_{\theta^{i}}\phi')-\int_{R} d^4xd^3\theta W(\theta') \mathcal{L}(\phi,\partial_{\mu}\phi,\partial_{\theta^{i}}\phi)\\
&=& \int_{R} d^4xd^3\theta W(\theta)\delta\mathcal{L}+\epsilon\int_{R} d^4xd^3\theta W(\theta)(\partial_{\mu}\delta x^{\mu})\mathcal{L}+\epsilon\int_{R} d^4xd^3\theta \partial_{\theta^{i}}(W(\theta)\delta \theta^{i})\mathcal{L} \label{varaction1}
\eea
Next we use the functional variance of field $\delta \phi(x,\theta)=\bar{\delta}\phi(x,\theta)+\delta x^{\mu}(\partial_{\mu}\phi)+\delta \theta^{i}(\partial_{\theta^{i}}\phi)$ and get,
\begin{multline}
\delta S=\int d^4xd^3\theta \Big\{ W\frac{\partial L}{\partial \phi}-W\partial_{\mu}\Big(\frac{\partial L}{\partial(\partial_{\mu}\phi)}\Big)-\partial_{\theta^{i}}\Big(W\frac{\partial L}{\partial (\partial_{\theta^{i}}\phi)}\Big)\Big\}\bar{\delta}\phi\\
+\int d^4xd^3\theta\Big[\partial_{\mu}(W\mathcal{L}\delta x^{\mu}+W\frac{\partial \mathcal{L}}{\partial(\partial_{\mu}\phi)}\bar{\delta}\phi)+\partial_{\theta^{i}}(W\mathcal{L}\delta \theta^{i}+W\frac{\partial \mathcal{L}}{\partial(\partial_{\theta^{i}}\phi)}\bar{\delta \phi})\Big] \label{varaction2}
\end{multline}
where $\bar{\delta} \phi=\phi'(x,\theta)-\phi(x,\theta)$.
After using Gauss divergence theorem and setting total derivative terms to vanish at the boundary $\partial R$, also by assuming a stationary value for S for an arbitrary variation of $\bar{\delta}\phi$ that vanishes on the boundary $\partial R$, we get Euler-Lagrange equation as,
\be
W\partial_{\mu}\Big(\frac{\partial \mathcal{L}}{\partial(\partial_{\mu}\phi)}\Big)-W\frac{\partial \mathcal{L}}{\partial \phi}+\partial_{\theta^{i}}\Big(W\frac{\partial \mathcal{L}}{\partial(\partial_{\theta^{i}}\phi)}\Big) =0. \label{gen-eom}
\ee
From surface terms in Eqn.(\ref{varaction2}), using Noether's theorem, we get conserved current $(J^{\mu},J^{\theta^{i}})$ as,
\be
J^{\mu}=W\Big \{\mathcal{L}\delta x^{\mu}+\Big(\frac{\partial \mathcal{L}}{\partial(\partial_{\mu}\phi)}\Big)\bar{\delta}\phi\Big \}=W\tilde{J}^{\mu} \label{current1}
\ee
and
\be
J^{\theta^{i}}=W\Big \{\mathcal{L}\delta \theta^{i}+\frac{\partial \mathcal{L}}{\partial(\partial_{\theta^{i}}\phi)}\bar{\delta}\phi) \Big\}=W\tilde{J}^{\theta^{i}} \label{current2}
\ee
such that
\be
\Omega=\partial_{\mu}(W\tilde{J}^{\mu})+\partial_{\theta^{i}}(W\tilde{J}^{\theta^{i}}) = 0. 
\ee
Now using the relation $\delta \phi(x,\theta)=\bar{\delta}\phi(x,\theta)+\delta x^{\mu}(\partial_{\mu}\phi)+\delta \theta^{i}(\partial_{\theta^{i}}\phi)$ in Eqn.(\ref{current1}) and Eqn.(\ref{current2}) eventually we get the components of the Energy-Momentum tensor as,
\bea
T^{\mu \nu}&=& \Big(\frac{\partial \mathcal{L}}{\partial (\partial_{\mu}\phi)}\Big)\partial^{\nu}\phi-\eta^{\mu \nu}\mathcal{L} \label{stressten} \\ 
T^{\theta^{i} \theta^{j}} &=& \Big(\frac{\partial \mathcal{L}}{\partial (\partial_{\theta^{i}}\phi)}\Big)\partial^{\theta^{j}}\phi-g^{\theta^{i} \theta^{j}}\mathcal{L} \\
T^{\mu \theta^{i}} &=& \Big(\frac{\partial \mathcal{L}}{\partial(\partial_{\mu}\phi)}\Big)\partial^{\theta^{i}}\phi \\
T^{\theta^{i}\mu}&=& \Big(\frac{\partial \mathcal{L}}{\partial(\partial_{\theta^{i}}\phi)}\Big)\partial^{\mu} \phi
\eea
Thus we get all the components of stress tensor for real scalar field theory in $7$-dimensional DFR space-time. Note that the Casimir effect will be calculated by taking the difference between the vacuum expectation value of the Energy-Momentum tensor at the right side of the plate and the left side of the plate. Also, note here that current densities and components of the stress tensor go over to the commutative results when $\theta=0$.
For constant value of weight function, (say $W(\theta)=1$) the Euler-Lagrange equation of motion in Eqn.(\ref{gen-eom}), current densities in Eqn.(\ref{current1}) and Eqn.(\ref{current2}), reduces to the results for DFRA real scalar field theory considered in \cite{amo5}. 
\section{Casimir Effect using DFRA scalar field }
In this section, we derive the Casimir force felt by one of the parallel plates by analyzing the vacuum fluctuation of the DFRA scalar field. For this, we first calculate the vacuum expectation value of the Energy-Momentum tensor corresponding to the DFRA scalar field theory, using the relation between the expectation value of the time-ordered product of fields and Green's function. Here the Green's function is obtained by solving Euler-Lagrange equations. Using this vacuum expectation value of Energy-Momentum tensor, we obtain corrections to Casimir force and Casimir energy. We investigate Casimir effect in $4+1$ dimensions $(x^{\mu},\theta)$ and the result we obtain is generalized to $4+3$-dimensional DFR space-time. Here the metric used is $diag(+------)$. 

We start with the DFRA massless real scalar field Lagrangian \cite{amo1,amo2,amo3,amo5},
\be
\mathcal{L}_0=\frac{1}{2}\partial_{\mu}\phi\partial^{\mu}\phi+\frac{\lambda^2}{2}\partial_{\theta^i}\phi\partial^{\theta^i}\phi \label{kg}
\ee
where in the limit $\lambda \rightarrow 0$, we get back Lagrangian of the real scalar field theory in commutative space-time.

To study Casimir effect between two parallel plates, we introduces these plates perpendicular to $x$-direction and model these plates, using delta function potentials at $x=0$ and $x=L$. Thus we get total Lagrangian as \cite{Mil}, 
\be
\mathcal{L} =\frac{1}{2}\partial_{\mu}\phi\partial^{\mu}\phi+\frac{\lambda^2}{2}\partial_{\theta^i}\phi\partial^{\theta^i}\phi -{\frac{\gamma}{2L}\phi^2\delta(x)}-{\frac{\gamma^\prime}{2L}}\phi^2\delta(x-L), \label{totint}
\ee
where $\gamma,\gamma'$ are coupling constant, which are dimension free quantities. Here we consider the Lagrangian in $4+1$ dimension given by
\be
\mathcal{L}=\frac{1}{2}{\partial_0\phi\partial_0\phi}-\frac{1}{2}{\partial_m\phi\partial_m\phi}+\frac{\lambda^2}{2}\partial_{\theta^1}\phi\partial^{\theta^1}\phi -{\frac{\gamma}{2L}\phi^2\delta(x)}-{\frac{\gamma^\prime}{2L}}\phi^2\delta(x-L); ~~m=1,2,3. \label{lag}
\ee
Note that the interaction part, mimicking the plates, is similar to that in the commutative space-time.

Using Eqn.(\ref{gen-eom}), we obtain Euler-Lagrange equation from the above Lagrangian as
\be
\Big(\partial_0^2 -\partial_m^2 -\lambda^2\partial_{\theta^{1}}^2+\frac{ \theta^{1}}{2\lambda^2}\partial_{\theta^{1}}+\frac{\gamma}{L}\delta(x)+\frac{\gamma'}{L}\delta(x-L)\Big)\phi({\bf x},\frac{\theta^{1}}{\lambda})=0, \label{eom}
\ee
where ${\bf x}$ denote all the three space co-ordinates $x,y,z$ and $\theta^{1}$ is the non-commutative of space coordinate.
Note that if weight function $W(\theta)$ is taken to be a constant, the third term, $\frac{ \theta^{1}}{2\lambda^2}\partial_{\theta^{1}}$ will be absent.
We re-express Eqn.(\ref{eom}) as,
\be
\Big(\partial_0^2-\partial_x^2-\partial_y^2-\partial_z^2-\lambda^2\partial_{\theta^{1}}^2 + \frac{ \theta^{1}}{2\lambda^2}\partial_{\theta^{1}}+\frac{\gamma}{L}\delta(x)+\frac{\gamma'}{L}\delta(x-L)\Big)\phi({\bf x},\frac{\theta^{1}}{\lambda})=0. \label{eom1}
\ee
Here in the limit, $\lambda \rightarrow 0$ and $\theta = 0$, we get back equation of motion in the commutative space-time \cite{Mil}. As we are considering only one non-commutative direction $\theta^{1}$, we set $\theta^{1}=\theta$.

Casimir force between two parallel plates is obtained by taking the difference between vacuum expectation value of $xx$ component of Energy-Momentum tensor at either side of the plate.
Since the vacuum expectation values of stress tensor is related to the Green's function as in Eqn.(\ref{ten-green}), first we calculate the Green's functions corresponding to Eqn.(\ref{eom1}). We start with,
\be
\Big(\partial_0^2-\partial_x^2-\partial_y^2-\partial_z^2-\lambda^2\partial_{\theta}^2 + \frac{\theta}{2\lambda^2}\partial_{\theta}+\frac{\gamma}{L}\delta(x)+\frac{\gamma'}{L}\delta(x-L)\Big)G({\bf x,x'};\frac{\theta}{\lambda},\frac{\theta'}{\lambda})=\delta^3({\bf x-x'})\delta(t-t')\delta(\frac{\theta}{\lambda}-\frac{\theta'}{\lambda}). \label{eom2}
\ee
Note that in the commutative limit, i.e., $\lambda \rightarrow 0$, $\theta =0$ above equation reduce to well known commutative result \cite{Mil}. In the commutative space-time one solves for the Green's function by using Fourier-transform for $t$, $y$ and $z$ dependence of the Green function. We can now treat the additional $\theta$-direction introduced by in the DFR space-time in the same manner as $y$ and $z$ directions. Then following the standard calculation we would get the Casimir force to be $F=-\frac{3.12}{32\pi^2L^5}$ (for weight function $W(\theta)=1$) which exactly match with the result obtained in \cite{Mil} for 5-dimensional commutative space-time. This shows that the newly introduced $\theta$-dimension act just like extra dimensions in commutative space-time. Note that this expression for the Casimir force is independent of $\theta$ and/or $\lambda$ and the Casimir force in DFR space-time does not have a commuatative limit; thus this Casimir effect does not distinguise between 5-dimensional DFR space-time and 5-dimensional commutative space-time. Since non-commutativity is expected to bring in modifications which should smoothly vanish in the commutative limit, we adopt a different calculation scheme in the following. In \cite{kam,prd79,plb668,farina,teo} extra dimensions were treated as compact directions and the Casimir effect was studied. We next study the Casimir effect in DFR space-time by treating additional dimensions introduced by non-commutativity as compact dimensions.

\subsection{Green's function solutions in different regions}
In this subsection, to study the Casimir effect in presence of extra compactified dimension, we compactify the extra $\theta$ - direction which comes due to the presence of non-commutativity of space-time. We then find the Green's function solutions in different regions in presence of this extra compactified dimension. Using Fourier transform, we rewrite Green's function in Eqn.(\ref{eom2}) as,
\be
G({\bf x,x'},t,t',\tilde{\theta},\tilde{\theta}')=\int_{-\infty}^{\infty} \frac{dw}{2\pi}\frac{dp_{y}}{2\pi}\frac{dp_{z}}{2\pi}e^{-i\omega(t-t')}e^{ip_{y}(y-y')}e^{ip_{z}(z-z')}g(x,x',\tilde{\theta},\tilde{\theta'},\omega,p_{\Vert}). \label{greenf}
\ee 
where $\frac{\theta}{\lambda}=\tilde{\theta}$, which has dimension of length, $p_{\Vert}=\sqrt{p_{y}^2+p_{z}^2}$ 
and $g(x,x',\tilde{\theta},\tilde{\theta'},\omega,p_{\Vert})$ is the reduced Green's function. Note here that we are not considering non-commutative coordinate $\theta$ in same footing as the other two transverse space dimensions, $y$ and $z$ but treats $\theta$ as a compactified direction. As discussed after Eqn.(\ref{eom2}), accounting $\theta$ coordinate in a way similar that of the transverse dimensions, leads to results which does not have proper commutative limit. To overcome this shortcoming here we will compactify this extra space dimension $\theta$ and study Casimir effect due to the presence of this extra compactified space dimension \cite{prd79,plb668,farina}. The $g(x,x',\tilde{\theta},\tilde{\theta'},\omega,p_{\Vert})$ obeys the equation,
\be
\Big({-\frac{\partial^2}{\partial x^2}}-\omega^2 + p_{\Vert}^2-\frac{\partial}{\partial \tilde{\theta}^2}+\frac{\tilde{\theta}}{2 \lambda^2}\frac{\partial}{\partial {\tilde{\theta}}}+\frac{\gamma}{L}\delta(x)+\frac{\gamma'}{L}\delta(x-L)\Big)g(x,x';\tilde{\theta},\tilde{\theta};\omega ; p_{\Vert})=\delta(x-x')\delta(\tilde{\theta}-\tilde{\theta}').\label{eom3}
\ee
where $p_{\Vert}^2=p_{y}^2+p_{z}^2$ and second term, $-\omega^2$ in above is coming due to action of $\partial_{0}^2$ over Green's function given in Eqn.(\ref{greenf}). Next we compactify $\tilde{\theta}$ direction by the definition $\frac{\tilde{\theta}}{R}=\eta$. Here $R$ is the size of the extra compactified dimension. Thus we use $\eta$ as the extra compactified direction, which varies from $0 \rightarrow 2\pi$ and we rewrite Eqn.(\ref{eom3}) as
\be
\Big({-\frac{\partial^2}{\partial x^2}}-\omega^2 + p_{\Vert}^2-\frac{1}{R^2}\frac{\partial}{\partial \eta^2}+\frac{\eta}{2 \lambda^2}\frac{\partial}{\partial {\eta}}+\frac{\gamma}{L}\delta(x)+\frac{\gamma'}{L}\delta(x-L)\Big)g(x,x';\eta,\eta';\omega ; p_{\Vert})=\delta(x-x')\sum_{n \in \mathbb{Z}}\delta(\eta-\eta'-n).\label{eom5}
\ee 
In above equation the reduced green's function is  
\be
g(x,x';\eta,\eta';\omega ; p_{\Vert})=\sum_{n \in \mathbb{Z}}g(x,x';\omega ; p_{\Vert},n)e^{i 2\pi n(\eta-\eta')} \label{eom6}
\ee 
and we use Poisson summation formula, $\sum_{n \in \mathbb{Z}}\delta(\eta-\eta'-n)=\sum_{n \in \mathbb{Z}}e^{i 2\pi n(\eta-\eta')}$. Thus we reexpress Eqn.(\ref{eom5}) as
\be
\sum_{n \in \mathbb{Z}}\bigg(\Big({-\frac{\partial^2}{\partial x^2}}-\omega^2 + p_{\Vert}^2+\frac{(2 \pi n)^2}{R^2}+\frac{\eta}{2 \lambda^2}(i 2 \pi n)+\frac{\gamma}{L}\delta(x)+\frac{\gamma'}{L}\delta(x-L)\Big)g(x,x';\omega ; p_{\Vert};n)-\delta(x-x')\bigg)e^{i 2\pi n(\eta-\eta')}=0.\label{eom7}
\ee
from which we find
\be
\Big({-\frac{\partial^2}{\partial x^2}}-\omega^2 + p_{\Vert}^2+C_{n}^2+\frac{\gamma}{L}\delta(x)+\frac{\gamma'}{L}\delta(x-L)\Big)g(x,x';\omega ; p_{\Vert};n)=\delta(x-x'),\label{eom8}
\ee
where $C_{n}^2=\frac{(2 \pi n)^2}{R^2}+\frac{\eta}{2 \lambda^2}(i 2 \pi n)$, which is a complex quantity. Here $n \in \mathbb{Z}$ and $\eta =\frac{\tilde{\theta}}{R}$. Next, we solve Eqn.(\ref{eom8}) using Dritchlet boundary condition and obtain the reduced Green's function in three regions as follows.

\bigskip

For region $x,x'<0$,
\begin{multline}
g = \frac{1}{2q}e^{-q|x-x'|}+{\frac{1}{2q\Delta}e^{q(x+x')}}\bigg({-{\frac{\gamma'}{2qL}\Big(1-\frac{\gamma}{2qL}\Big)}{-{\frac{\gamma}{2qL}}\Big(1+\frac{\gamma'}{2qL}\Big)e^{2qL}}}\bigg)\\ 
+ \sum_{n \in \mathbb{Z} \backslash \left\lbrace 0 \right\rbrace} \Big [\frac{1}{2\tilde{q}}e^{-\tilde{q}|x-x'|}+{\frac{1}{2\tilde{q}\tilde{\Delta}}e^{\tilde{q}(x+x')}}\bigg({-{\frac{\gamma'}{2\tilde{q}L}\Big(1-\frac{\gamma}{2\tilde{q}L}\Big)}{-{\frac{\gamma}{2\tilde{q}L}}\Big(1+\frac{\gamma'}{2\tilde{q}L}\Big)e^{2\tilde{q}L}}}\bigg)\Big]e^{i 2 \pi n (\eta-\eta')}; \label{reg1} 
\end{multline}

for region $0<x,x'<L$,
\begin{multline}
g=\frac{1}{2q}e^{-q|x-x'|}+\frac{1}{2q\Delta}\bigg(\frac{\gamma\gamma'}{(2qL)^2}2\textnormal{cosh}(q|x-x'|)-{\frac{\gamma}{2qL}\Big(1+\frac{\gamma'}{2qL}\Big)e^{2qL}e^{-q(x+x')}}-{\frac{\gamma'}{2qL}\Big(1+\frac{\gamma}{2qL}\Big)e^{q(x+x')}}\bigg)\\
+ \sum_{n \in \mathbb{Z} \backslash \left\lbrace 0 \right\rbrace}\Big[\frac{1}{2\tilde{q}}e^{-\tilde{q}|x-x'|}+\frac{1}{2\tilde{q}\tilde{\Delta}}\bigg(\frac{\gamma\gamma'}{(2\tilde{q}L)^2}2\textnormal{cosh}(|\tilde{q}x-x'|)-{\frac{\gamma}{2\tilde{q}L}\Big(1+\frac{\gamma'}{2\tilde{q}L}\Big)e^{2\tilde{q}L}e^{-\tilde{q}(x+x')}} \\-{\frac{\gamma'}{2\tilde{q}L}\Big(1+\frac{\gamma}{2\tilde{q}L}\Big)e^{\tilde{q}(x+x')}}\bigg)\Big] e^{i 2 \pi n (\eta-\eta')};  \label{reg2}
\end{multline}

for region $L<x,x'$,
\begin{multline}
g =\frac{1}{2q}e^{-q|x-x'|}+{\frac{1}{2q\Delta}e^{-q(x+x'-2L)}}\bigg({-{\frac{\gamma}{2qL}\Big(1-\frac{\gamma'}{2qL}\Big)}}{-{\frac{\gamma'}{2qL}\Big(1+\frac{\gamma}{2qL}}\Big)e^{2qL}}\bigg)\\
+\sum_{n \in \mathbb{Z} \backslash \left\lbrace 0 \right\rbrace}\Big[\frac{1}{2\tilde{q}}e^{-\tilde{q}|x-x'|}+{\frac{1}{2\tilde{q}\tilde{\Delta}}e^{-\tilde{q}(x+x'-2L)}}\bigg({-{\frac{\gamma}{2\tilde{q}L}\Big(1-\frac{\gamma'}{2\tilde{q}L}\Big)}}{-{\frac{\gamma'}{2\tilde{q}L}\Big(1+\frac{\gamma}{2\tilde{q}L}}\Big)e^{2\tilde{q}L}}\bigg)\Big]e^{i 2 \pi n (\eta-\eta')}; \label{reg3}
\end{multline}
Here, ${n \in \mathbb{Z} \backslash \left\lbrace 0 \right\rbrace}$ imply that $n$ can take any integer value except 0.
In the above equations we have defined $q=-\omega^2+p_{\Vert}^2$ and $\tilde{q}=\sqrt{C_{n}^2-\omega^2+p_{\Vert}^2}$,where $C_{n}^2=\frac{(2 \pi n)^2}{R^2}+\frac{\eta}{2 \lambda^2}(i 2 \pi n)$. And we have also used
\be 
\Delta=\Big(1+\frac{\gamma}{2qL}\Big)\Big(1+\frac{\gamma'}{2qL}\Big)e^{2qL}-\frac{\gamma\gamma'}{(2qL)^2}\label{del}
\ee
and 
\be 
\tilde{\Delta}=\Big(1+\frac{\gamma}{2\tilde{q}L}\Big)\Big(1+\frac{\gamma'}{2\tilde{q}L}\Big)e^{2\tilde{q}L}-\frac{\gamma\gamma'}{(2\tilde{q}L)^2}.\label{del1}
\ee
Note here that in the limit $R \rightarrow 0$ (size of the compactified dimension), $\tilde{q} \rightarrow \infty$ and we get commutative results of the reduced Green's function solutions \cite{Mil} for all the three regions mentioned above.

\subsection{Modification of Energy-Momentum tensor}
In this subsection, we obtain the stress tensor for DFRA scalar field theory in $4+1$ dimensions using the Lagrangian given in Eqn.(\ref{lag}) in Eqn.(\ref{stressten}). Thus we find the stress tensor as
\bea 
T^{\mu \nu} & = & \partial^{\mu}\phi\partial^{\nu}\phi -\eta^{\mu\nu}\mathcal{L} \nonumber \\
& = & \partial^{\mu}\phi\partial^{\nu}\phi -\eta^{\mu\nu}\frac{1}{2}\Big(\partial^{0}\phi\partial^{0}\phi -\partial^{m}\phi\partial^{m}\phi+\partial_{\tilde{\theta}}\phi\partial^{\tilde{\theta}}\phi-\frac{\gamma}{L}\delta(x)-\frac{\gamma'}{L}\delta(x-L)\Big) \label{st}
\eea
In the limit $\lambda \rightarrow 0$ and $\theta=0$, above $T^{\mu\nu}$ reduces to that of the real scalar field in the commutative space-time. We re-express Energy-Momentum tensor as
\be
\hat{T}^{\mu\nu}_{{\bf x},{\bf x'},\theta,\theta'}=\frac{1}{2}\bigg((\partial^{\mu} \partial^{' \nu}+\partial^{\mu}\partial^{' \nu})-\eta^{\mu\nu}\Big(\partial_0\partial^{'}_0-\partial_m\partial^{'}_m-\partial_{\tilde{\theta}}\partial^{'}_{\tilde{\theta}}-\frac{\gamma}{L}\delta(0)-\frac{\gamma'}{L}\delta(x-L)\Big)\bigg)\phi({\bf x},\tilde{\theta})\phi({\bf x'},\tilde{\theta}') \label{st1}
\ee
where $\tilde{\theta}=\frac{\theta}{\lambda}$ and has dimension of $L$. We express this Energy-Momentum tensor as an operator acting on the time-ordered product of DFRA scalar fields, i.e.,
\be
\hat{T}^{\mu\nu}_{{\bf x},{\bf x'},\theta,\theta'}=\hat{O}_{\lambda,\theta}^{\mu \nu}(\partial, \partial^{'})T(\phi({\bf x},\tilde{\theta})\phi({\bf x'},\tilde{\theta}')), \label{st1}
\ee
where 
\be
\hat{O}_{\lambda,\theta}^{\mu \nu}(\partial, \partial^{'})=\frac{1}{2}\bigg((\partial^{\mu} \partial^{' \nu}+\partial^{\mu}\partial^{' \nu})-\eta^{\mu\nu}\Big(\partial_0\partial^{'}_0-\partial_m\partial^{'}_m-\partial_{\tilde{\theta}}\partial_{\tilde{\theta}}^{'}-\frac{\gamma}{L}\delta(x)-\frac{\gamma'}{L}\delta(x-L)\Big)\bigg). \label{operator}
\ee
Note here that the $\lambda$ and $\theta$ in the subscript of the $\hat{O}$ is to imply that operator has terms due to non-commutativity of space-time. As $\lambda \rightarrow 0$, $\theta=0$ we get back the commutative result.

Next we take vacuum expectation value of the Energy-Momentum tensor given in Eqn.(\ref{st1}) and find,
\bea
<\hat{T}^{\mu\nu}_{{\bf x},{\bf x'},\theta,\theta'}> &=& \hat{O}_{\lambda,\theta}^{\mu \nu}(\partial, \partial^{'})<T(\phi({\bf x},\tilde{\theta})\phi({\bf x'},\tilde{\theta}'))> \\ \nonumber
&=& -i \frac{1}{2}\bigg((\partial^{\mu} \partial^{' \nu}+\partial^{\mu}\partial^{' \nu})-\eta^{\mu\nu}\Big(\partial_0\partial^{'}_0-\partial_m\partial^{'}_m-\partial_{\tilde{\theta}}\partial_{\tilde{\theta}}^{'}\Big)\bigg)G_{\lambda,\theta}({\bf x,x'};\tilde{\theta},\tilde{\theta}')\label{vacst}
\eea
In the above, we have used the identity $i<T(\phi\phi')>=G$, where G is Green's function.

In the limit, $\lambda \rightarrow 0$ and $\theta=0$ we find $G({\bf x,x'};\tilde{\theta},\tilde{\theta}')$ reduces to $G({\bf x,x'})$-the Green's function for the commutative case. Now from above equation, we evaluate vacuum expectation value of $xx$ component of Energy-Momentum tensor as
\bea
<\hat{T}^{xx}>&=&-i\bigg(\frac{1}{2}\partial_0\partial'_{0}+\frac{1}{2}\partial_x\partial'_{x}-\frac{1}{2}\partial_y\partial'_{y}-\frac{1}{2}\partial_z\partial'_{z}-\frac{1}{2}\partial_{\tilde{\theta}}\partial_{\tilde{\theta}}^{'}\bigg)G({\bf x,x'};\tilde{\theta},\tilde{\theta}').\label{vacst1}
\eea
Next we compactify $\theta$ direction (see the discussion after Eqn.(\ref{eom3})) by defining $\eta=\frac{\tilde{\theta}}{R}$ and $\eta'=\frac{\tilde{\theta}'}{R}$, where $\eta$ and $\eta'$ varies from $0 \rightarrow 2\pi$ and $R$ is the size of the compact dimension $\eta$. Thus above equation becomes
\bea
<\hat{T}^{xx}>&=&-i\bigg(\frac{1}{2}\partial_0\partial'_{0}+\frac{1}{2}\partial_x\partial'_{x}-\frac{1}{2}\partial_y\partial'_{y}-\frac{1}{2}\partial_z\partial'_{z}-\frac{1}{2R^2}\partial_{\eta}\partial_{\eta}^{'}\bigg)G({\bf x,x'};\eta,\eta'),\label{vacst1}
\eea
Next we use $G$ given in Eqn.(\ref{greenf}) and the reduced Energy-Momentum tensor defined through
\be
<\hat{T}^{\mu \nu}>=\int \frac{d\omega}{2 \pi}\frac{dp_{y}}{2\pi}\frac{dp_{z}}{2\pi}\hat{t}^{\mu\nu} \label{redst}
\ee
to obtain $xx$ component of reduced stress tensor
\be
\hat{t}_{xx}=\frac{1}{2i}\bigg(\omega^2-p_{\Vert}^2+\partial_x\partial'_{x}-\frac{1}{R^2}\partial_{\eta}\partial_{\eta}^{'}\bigg)g(x,x^{'};\eta,\eta';\omega ;p_{\Vert})\Big|_{x=x';\eta=\eta'} \label{redst1}
\ee
Thus we get $ xx$ component of reduced stress tensor for DFRA scalar field in presence of compactified dimension coming due to presence of non-commutativity in space-time. 
\subsection{Modified Casimir Force and Casimir Energy }
In this subsection, we evaluate the Casimir force experienced by the plate at $x=L$ and corresponding Casimir energy. This is calculated using the discontinuity between the component of stress tensors (or pressures) acting on it from the left side of the plate and the right side of the plate. For this, we first find the reduced energy-momentum tensors $t_{xx}\Big|_{x=L^{-}}$ and $t_{xx}\Big|_{x=L+}$ and by taking the difference between these two we get the Casimir force.

Now using Eqn.(\ref{reg2}) in Eqn.(\ref{redst1}), we get pressure acting from the left side on the plate at $x=L$ as
\begin{multline}
\hat{t}_{xx}\Big |_{x=L^{-}} = \frac{1}{2i}\bigg[\Big(-q-\frac{q}{\Delta}\frac{2\gamma\gamma'}{(2qL)^2}\Big) + \sum_{n \in \mathbb{Z} \backslash \left\lbrace 0 \right\rbrace}\Big(-\tilde{q}-\frac{\tilde{q}}{\tilde{\Delta}}\frac{2\gamma\gamma'}{(2\tilde{q}L)^2}\Big) \\ + \sum_{n \in \mathbb{Z} \backslash \left\lbrace 0 \right\rbrace}\bigg(C_{n}^2+{\Big(\frac{2\pi n}{R}\Big)}^2\bigg)\bigg( \frac{1}{2\tilde{q}}+\frac{1}{2\tilde{q}\tilde{\Delta}} \Big\{\frac{\gamma \gamma'}{(2\tilde{q}L)^2}-\frac{\gamma}{2\tilde{q}L}\Big(1+\frac{\gamma'}{2\tilde{q}L} \Big)e^{2\tilde{q}(L-x)}-\frac{\gamma'}{2\tilde{q}L}\Big(1+\frac{\gamma}{2\tilde{q}L} \Big)e^{2\tilde{q}x} \Big\} \bigg) \bigg]. \label{redstleft2}
\end{multline}
Similarly by substituting Eqn.(\ref{reg3}) in Eqn.(\ref{redst1}), we find the $xx$ component of reduced Energy-Momentum tensor, just to the right of the plate,(i.e., pressure acting from the right side) as 
\begin{multline}
\hat{t}_{xx}\Big |_{x=L^{+}} = \frac{1}{2i}\bigg[(-q)+ \sum_{n \in \mathbb{Z} \backslash \left\lbrace 0 \right\rbrace}(-\tilde{q}) \\ + \sum_{n \in \mathbb{Z} \backslash \left\lbrace 0 \right\rbrace}\bigg(C_{n}^2+{\Big(\frac{2\pi n}{R}\Big)}^2\bigg)\bigg( \frac{1}{2\tilde{q}}+\frac{1}{2\tilde{q}\tilde{\Delta}}e^{2\tilde{q}(L-x)} \Big\{-\frac{\gamma'}{2\tilde{q}L}\Big(1+\frac{\gamma}{2\tilde{q}L}\Big)e^{2\tilde{q}L}-\frac{\gamma}{2\tilde{q}L}\Big(1-\frac{\gamma'}{2\tilde{q}L} \Big) \Big\} \bigg) \bigg]. \label{redstright}
\end{multline}
In the above equation, $q^2=-\omega^2+p_{\Vert}^2$ and $\tilde{q}=\sqrt{C_{n}^2-\omega^2+p_{\Vert}^2}$, where $p_{\Vert}^2=p_{y}^2+p_{z}^2$.

Now Casimir force arising due to the vacuum fluctuation, acting on the plate at $x=L$ is
\bea
F &=& <\hat{T}^{xx}>\Big |_{x=L^{-}}-<\hat{T}^{xx}>\Big |_{x=L^{+}} \nonumber \\
&=& \int \frac{d\omega}{2 \pi}\frac{dp_{y}}{2\pi}\frac{dp_{z}}{2\pi}\Big(\hat{t}^{xx}\Big |_{x=L^{-}}-\hat{t}^{xx}\Big |_{x=L^{+}}\Big), \label{force}
\eea
Using Eqn.(\ref{redstleft2}) and Eqn.(\ref{redstright}) in above equation we get Casimir force as
\be
F= \frac{1}{2i}\bigg[\int \frac{d\omega}{2 \pi}\frac{dp_{y}}{2\pi}\frac{dp_{z}}{2\pi}\Big\{\frac{-2q}{\Delta}\frac{\gamma \gamma'}{(2qL)^2}+\sum_{n \in \mathbb{Z} \backslash \left\lbrace 0 \right\rbrace}\bigg(\frac{-2\tilde{q}}{\tilde{\Delta}}\frac{\gamma \gamma'}{(2\tilde{q}L)^2}\bigg)+\sum_{n \in \mathbb{Z} \backslash \left\lbrace 0 \right\rbrace}\bigg(C_{n}^2+{\Big(\frac{2\pi n}{R}\Big)}^2\bigg)\frac{1}{2\tilde{q}\tilde{\Delta}}\frac{\gamma \gamma'}{(2\tilde{q}L)^2}\Big\}\bigg]. \label{force1}
\ee
where $\tilde{q}=\sqrt{C_{n}^2-\omega^2+p_{y}^2+p_{z}^2}$. After rewriting the frequency as $\omega=i\zeta$ and using relations given in Eqn.(\ref{del}) and Eqn.(\ref{del1}),  after straightforward simplifications we get 

\begin{multline} 
F = -\bigg[\int \frac{d\zeta}{4 \pi}\frac{dp_{y}}{2\pi}\frac{dp_{z}}{2\pi}\Big\{\frac{2q}{\Big(\frac{2qL}{\gamma}+1\Big)\Big(\frac{2qL}{\gamma'}+1\Big)e^{2qL}-1} + \sum_{n \in \mathbb{Z} \backslash \left\lbrace 0 \right\rbrace}\frac{2\tilde{q}}{\Big(\frac{2\tilde{q}L}{\gamma}+1\Big)\Big(\frac{2\tilde{q}L}{\gamma'}+1\Big)e^{2\tilde{q}L}-1}\\ -\sum_{n \in \mathbb{Z} \backslash \left\lbrace 0 \right\rbrace}\bigg(C_{n}^2+{\Big(\frac{2\pi n}{R}\Big)}^2\bigg)\frac{1}{2\tilde{q}\Big(\Big(\frac{2\tilde{q}L}{\gamma}+1\Big)\Big(\frac{2\tilde{q}L}{\gamma'}+1\Big)e^{2\tilde{q}L}-1\Big)} \Big\}\bigg]. \label{force3}
\end{multline}
Here we take strong interaction limit, i.e., $\gamma,\gamma' \rightarrow \infty $ and we get
\be
F_{\gamma,\gamma' \rightarrow \infty} = -\bigg[\int \frac{d\zeta}{4 \pi}\frac{dp_{y}}{2\pi}\frac{dp_{z}}{2\pi}\Big\{\frac{2q}{e^{2qL}-1} + \sum_{n \in \mathbb{Z} \backslash \left\lbrace 0 \right\rbrace}\frac{2\tilde{q}}{e^{2\tilde{q}L}-1}-\sum_{n \in \mathbb{Z} \backslash \left\lbrace 0 \right\rbrace}\bigg(C_{n}^2+{\Big(\frac{2\pi n}{R}\Big)}^2\bigg)\frac{1}{2\tilde{q}\Big(e^{2\tilde{q}L}-1\Big)} \Big\}\bigg]. \label{force4}
\ee
Using relations $P=\sqrt{\zeta^2+p_{y}^2+p_{z}^2}$ and $d\zeta dp_{y} dp_{z}=4\pi P^2 dP$ (after integration of angular variable), the above equation is re-written as
\be
F_{\gamma,\gamma' \rightarrow \infty}= -\bigg[\int_{0}^{\infty} \frac{P^2dP}{4 \pi^2}\Big\{\frac{2q}{e^{2qL}-1} + \sum_{n \in \mathbb{Z} \backslash \left\lbrace 0 \right\rbrace}\frac{2\tilde{q}}{e^{2\tilde{q}L}-1}-\sum_{n \in \mathbb{Z} \backslash \left\lbrace 0 \right\rbrace}\bigg(C_{n}^2+{\Big(\frac{2\pi n}{R}\Big)}^2\bigg)\frac{1}{2\tilde{q}\Big(e^{2\tilde{q}L}-1\Big)} \Big\}\bigg]\label{force5}
\ee
where now $q=P$ and $\tilde{q}=\sqrt{C_{n}^2+P^2}$. We rewrite the above equation as
\begin{multline}
F_{\gamma,\gamma' \rightarrow \infty}= -\bigg[\int_{0}^{\infty} \frac{(2PL)^2d(2PL)}{32 \pi^2 L^4}\Big\{\frac{2PL}{(e^{2PL}-1)} + \sum_{n \in \mathbb{Z} \backslash \left\lbrace 0 \right\rbrace}\frac{\sqrt{4C_{n}^2L^2+4P^2L^2}}{(e^{\sqrt{4C_{n}^2L^2+4P^2L^2}}-1)} \\ -\sum_{n \in \mathbb{Z} \backslash \left\lbrace 0 \right\rbrace}\bigg(C_{n}^2+{\Big(\frac{2\pi n}{R}\Big)}^2\bigg)\frac{L^2}{\sqrt{4C_{n}^2L^2+4P^2L^2}\Big(e^{\sqrt{4C_{n}^2L^2+4P^2L^2}}-1\Big)} \Big\}\bigg]. \label{force6}
\end{multline}
Note that the $C_{n}^2$-dependent terms are all due to the non-commutativity of the space-time (see Eqn.(\ref{eom8})). Next we rewrite the above equation as
\begin{multline}
F_{\gamma,\gamma' \rightarrow \infty}= -\frac{1}{32\pi^2 L^4}\int_{0}^{\infty} \bigg \{ \frac{Y^3dY}{(e^{Y}-1)} + \sum_{n \in \mathbb{Z} \backslash \left\lbrace 0 \right\rbrace}\frac{Y^2\sqrt{4C_{n}^2L^2+Y^2}dY}{(e^{\sqrt{4C_{n}^2L^2+Y^2}}-1)} \\ -\sum_{n \in \mathbb{Z} \backslash \left\lbrace 0 \right\rbrace}\bigg(C_{n}^2+{\Big(\frac{2\pi n}{R}\Big)}^2\bigg)\frac{Y^2L^2dY}{\sqrt{4C_{n}^2L^2+Y^2}\Big(e^{\sqrt{4C_{n}^2L^2+Y^2}}-1\Big)} \bigg\} \label{force7}
\end{multline}
where we use $Y=2PL$. After carrying out the integration, we obtain the modified Casimir force as
\be
F_{\gamma,\gamma' \rightarrow \infty}=-\bigg[\frac{\pi^2}{480L^4}+\sum_{n \in \mathbb{Z} \backslash \left\lbrace 0 \right\rbrace}\sum_{m=1}\bigg\{\Big(\frac{3}{32\pi^2 L^2}\Big)\Big(\frac{C_{n}}{m}\Big)^2K_{2}(2LmC_{n})+\frac{1}{16\pi^2 L}\bigg(3C_{n}^2-\Big(\frac{2\pi n}{R}\Big)^2\bigg)\frac{C_{n}}{m}K_{1}(2LmC_{n})\bigg\}\bigg]. \label{forcefinal1}
\ee
Note here that $C_{n}^2=\frac{(2 \pi n)^2}{R^2}+\frac{\eta}{2 \lambda^2}(i 2 \pi n)$ and  $C_{n}^2$ dependent term in the above equation is purely due to the non-commutativity of space-time. From expression of $C_{n}^2$ we can see that the Casimir force is complex, which is not expected. The complex part in the $C_{n}^2$ is coming due to presence of the weight function $W(\theta)$ in the theory (see Eqn.(\ref{action1})). $C_{n}$ can be made real by taking weigh function to be constant (say $W(\theta)=1)$ as any $\theta$ dependent form of weight function will make the Casimir force and energy complex. Thus by taking ($W(\theta)=1$) in our study we find the Casimir force to be
\be
F_{\gamma,\gamma' \rightarrow \infty}=-\bigg[\frac{\pi^2}{480L^4}+\sum_{n \in \mathbb{Z} \backslash \left\lbrace 0 \right\rbrace}\sum_{m=1}\bigg\{\Big(\frac{3}{8L^2R^2}\Big)\Big(\frac{n}{m}\Big)^2K_{2}\Big(\frac{4Lmn\pi}{R}\Big)+\frac{n^3\pi}{ mLR^3}K_{1}\Big(\frac{4Lmn\pi}{R}\Big)\bigg\}\bigg] \label{forcefinal2}
\ee
which is similar to the result obtained in \cite{farina} for the Casimir effect in presence of extra one compactified dimension. Here $K_{1}$ and $K_{2}$ are modified Bessel function of second kind, which have dependency on seperation length, $L$ and size of extra compactified dimension, $R$.

Note here that $R$ is the size of the compactified dimension due to presence of noncommutativity in space-time and in the limit $R \rightarrow 0$, above expression of Casimir force reduce to the commutative Casimir force expression \cite{Mil,kam}.

We also derive the modified Casimir energy as
\begin{multline}
E_{\gamma,\gamma'\rightarrow \infty} = -\int F_{\gamma,\gamma' \rightarrow \infty}dL\\
= -\frac{\pi^2}{1440L^3}-\sum_{n \in \mathbb{Z} \backslash \left\lbrace 0 \right\rbrace}\sum_{m=1}\bigg\{\Big(\frac{3}{32LR^2}\Big)\Big(\frac{n}{m}\Big)^2 G_{2,1}^{1,3}\left(\frac{4Lmn\pi}{R},\frac{1}{2}\middle\vert
\begin{array}{c}
\frac{3}{2}\\
-1,1,\frac{1}{2}
\end{array}
\right)\\ +
\frac{n^3\pi}{ 4mR^3}G_{2,1}^{1,3}\left(\frac{4Lmn\pi}{R},\frac{1}{2}\middle\vert
\begin{array}{c}
\frac{3}{2}\\
-\frac{1}{2},\frac{1}{2},0
\end{array}
\right)\bigg\}
\end{multline}
where in the above, $G_{2,1}^{1,3}$ is the Meijer-G funtion. Note here that Casimir energy has correction terms, $L$ dependent Meijer-G function whose coefficient vary as $\frac{1}{L}$. In the commutative limit, i.e., $R \rightarrow 0$, we get back the commutative result for the Casimir energy \cite{Mil,kam}.

Note that apart from terms proportional to $L^{-4}$, non-commutativity also bring terms that vary as $L^{-2}$ and $L^{-1}$ in the Casimir force (see Eqn.(\ref{forcefinal2})). And also Casimir force has correction terms that vary as $\frac{1}{R^2}$ and $\frac{1}{R^3}$, where $R$ is the size of the extra compactified dimension $\theta$ due to presence of non-commutativity in the space-time. 

\section{Thermal Correction to Casimir effect}
In this section, we study the finite temperature correction to the Casimir effect in DFR space-time and analyze the modifications to the Casimir effect at high temperature and low-temperature limits.
To obtain finite temperature correction to the Casimir force we substitute $\zeta =\frac{2\pi m}{\beta}$ \cite{kam,Brevik} in Eqn.(\ref{force4}). After this substitution, we replace the integration with summation as
\be
\int_{-\infty}^{\infty}\frac{d\zeta}{2\pi} ~ \rightarrow ~ \frac{2}{\beta}\sum_{m=0}^{\infty} \prime, ~~m ~\in ~Z \label{tempcas}
\ee
where $\prime$ implies the counting of $m=0$ term with $\frac{1}{2}$ factor and here $\beta=\frac{1}{KT}$, where K is the Boltzmann constant. Using this prescription, we re-express the force expression in Eqn.(\ref{force4}) as
\be
F_{\gamma,\gamma' \rightarrow \infty} = - \frac{2}{\beta}\sum_{m=0}^{\infty}{\prime}\int \frac{dp_{y}}{4\pi}\frac{dp_{z}}{2\pi}\bigg[\frac{2q}{e^{2qL}-1} + \sum_{n \in \mathbb{Z} \backslash \left\lbrace 0 \right\rbrace}\frac{2\tilde{q}}{e^{2\tilde{q}L}-1}-\sum_{n \in \mathbb{Z} \backslash \left\lbrace 0 \right\rbrace}\bigg(C_{n}^2+{\Big(\frac{2\pi n}{R}\Big)}^2\bigg)\frac{1}{2\tilde{q}\Big(e^{2\tilde{q}L}-1\Big)}\bigg], \label{tempcas1}
\ee 
where, $q=\sqrt{\zeta^2+p_{y}^2+p_{z}^2}$, $\tilde{q}=\sqrt{C_{n}^2+\zeta^2+p_{y}^2+p_{z}^2}$, and $C_{n}^2=\frac{(2 \pi n)^2}{R^2}$ (note that we have set $W(\theta)=1$). Note here that apart from the overall multiplication of $\frac{2}{\beta}$ in above equation, $\beta$ dependency is appearing through $\tilde{q}$ also. Next we use polar form of measure $dp_{y}dp_{z}=2\pi p_{\Vert}dp_{\Vert}$ where $p_{\Vert}=\sqrt{p_{y}^2 + p_{z}^2}$, re-define $dp_{\Vert}^2=dq^2$ and
\be
t= \frac{4\pi L}{\beta} \label{temp}
\ee
and re-express above equation as
\begin{multline}
F_{\gamma,\gamma' \rightarrow \infty} = -\frac{2}{\beta}\sum_{m=0}^{\infty}{\prime}\bigg[\int_{mt}^{\infty} \frac{Y^2dY}{16\pi L^3(e^{Y}-1)}+\sum_{n \in \mathbb{Z} \backslash \left\lbrace 0 \right\rbrace}\int_{\sqrt{(mt)^2+4L^2C_{n}^2}}^{\infty} \frac{\tilde{Y}^2d\tilde{Y}}{16\pi L^3(e^{\tilde{Y}}-1)} \\ - \sum_{n \in \mathbb{Z} \backslash \left\lbrace 0 \right\rbrace}C_{n}^2\int_{\sqrt{(mt)^2+4L^2C_{n}^2}}^{\infty}\frac{d\tilde{Y}}{8 \pi L(e^{\tilde{Y}}-1)}\bigg]. \label{tempcas2}
\end{multline}
Here $Y=2qL=\sqrt{(mt)^2 + 4L^2p_{\Vert}^2}$ and $\tilde{Y}=2\tilde{q}L=\sqrt{(mt)^2 +4L^2C_{n}^2 +4L^2p_{\Vert}^2}$. Note that through $Y$ and $\tilde{Y}$, dependence on t ( i.e., $\beta$ dependence) is entering the above expression for force (see Eqn.(\ref{temp})). We investigate the high temperature limit and the low temperature limit of the finite temperature correction of the Casimir effect as follows.

\bigskip
For high temperature limit: since from Eqn.(\ref{temp}) we get $t=4\pi LKT$ and thus for large value of T, we find $t=4\pi LKT>>1$ i.e., $4\pi L >> \beta$ \cite{kam,Brevik}. After evaluation of integration in above equation using zeta function, we get Casimir force in the high temperature limit as
\begin{multline}
F_{4 \pi L >> \beta}=-\bigg[\Big\{\frac{1}{8\pi \beta L^3}\zeta(3)+\frac{1}{4\pi\beta L^3}(1+t+\frac{t^2}{2})e^{-t} \Big\} + \frac{1}{16 \pi L^3 \beta}\sum_{n \in \mathbb{Z} \backslash \left\lbrace 0 \right\rbrace}\Big(4L^2C_{n}^2+4LC_{n}+2\Big)e^{-2LC_{n}} \\+\frac{1}{8\pi L^3\beta}\sum_{n \in \mathbb{Z} \backslash \left\lbrace 0 \right\rbrace}\Big((t^2+4L^2C_{n}^2)+2\sqrt{t^2+4L^2C_{n}^2}+2\Big)e^{-\sqrt{t^2+4L^2C_{n}^2}}  \\-\frac{1}{4\pi L\beta}\sum_{n \in \mathbb{Z} \backslash \left\lbrace 0 \right\rbrace}C_{n}^2\bigg(\frac{e^{-2LC_{n}}}{2}+e^{-\sqrt{t^2+4L^2C_{n}^2}}\bigg)\bigg]. \label{hightempcas}
\end{multline}
Note here that apart from the explicit $\beta$ dependency in all the terms, through t (see Eqn.(\ref{temp})) also , $\beta$ enter the above expression for force at the high-temperature limit . The above expression for the Casimir force for the high temperature limit with $C_{n}^2=\frac{(2 \pi n)^2}{R^2}$ reduce to commutative result \cite{kam,Brevik,Mil} in the limit $R \rightarrow 0$, where R is the size of the extra compactified dimension.

For low temperature limit: here we study the low temperature limit of Casimir effect using Poisson summation formula \cite{kam,Brevik,Mil};
any function $b(X)$ and it's Fourier transform $c(\alpha)$ i.e.,
\be
c(\alpha)=\frac{1}{2\pi}\int_{-\infty}^{\infty} b(X)e^{-i\alpha X} dX , \label{poisson1}
\ee
satisfy the identity
\be
\sum_{-\infty}^{\infty}b(m)=2\pi \sum_{-\infty}^{\infty}c(2\pi m). \label{poisson2}
\ee
Using this in Eqn.(\ref{tempcas2}) with
\begin{multline}
b(m)= \bigg[\int_{mt}^{\infty} \frac{Y^2dY}{16\pi L^3(e^{Y}-1)}+\sum_{n \in \mathbb{Z} \backslash \left\lbrace 0 \right\rbrace}\int_{\sqrt{(mt)^2+4L^2C_{n}^2}}^{\infty} \frac{\tilde{Y}^2d\tilde{Y}}{16\pi L^3(e^{\tilde{Y}}-1)} \\ - \sum_{n \in \mathbb{Z} \backslash \left\lbrace 0 \right\rbrace}C_{n}^2\int_{\sqrt{(mt)^2+4L^2C_{n}^2}}^{\infty}\frac{d\tilde{Y}}{8 \pi L(e^{\tilde{Y}}-1)}\bigg] \bigg|_{m=X}, \label{lowtemp1}
\end{multline}
where $Y=2qL=\sqrt{(mt)^2 + 4L^2p_{\Vert}^2}$ and $\tilde{Y}=2\tilde{q}L=\sqrt{(mt)^2 +4L^2C_{n}^2 +4L^2p_{\Vert}^2}$, we find
\be
\begin{split}
c(\alpha) = \frac{1}{2\pi}\int_{-\infty}^{\infty}e^{-i\alpha X}dX\bigg[\int_{Xt}^{\infty} \frac{Y^2dY}{16\pi L^3(e^{Y}-1)}+\sum_{n \in \mathbb{Z} \backslash \left\lbrace 0 \right\rbrace}\int_{\sqrt{(Xt)^2+4L^2C_{n}^2}}^{\infty} \frac{\tilde{Y}^2d\tilde{Y}}{16\pi L^3(e^{\tilde{Y}}-1)} \\ - \sum_{n \in \mathbb{Z} \backslash \left\lbrace 0 \right\rbrace}C_{n}^2\int_{\sqrt{(Xt)^2+4L^2C_{n}^2}}^{\infty}\frac{d\tilde{Y}}{8 \pi L(e^{\tilde{Y}}-1)}\bigg] \\
= \frac{1}{\pi}\int_{0}^{\infty}cos(\alpha X)dX\bigg[\int_{Xt}^{\infty} \frac{Y^2dY}{16\pi L^3(e^{Y}-1)}+\sum_{n \in \mathbb{Z} \backslash \left\lbrace 0 \right\rbrace}\int_{\sqrt{(Xt)^2+4L^2C_{n}^2}}^{\infty} \frac{\tilde{Y}^2d\tilde{Y}}{16\pi L^3(e^{\tilde{Y}}-1)} \\ - \sum_{n \in \mathbb{Z} \backslash \left\lbrace 0 \right\rbrace}C_{n}^2\int_{\sqrt{(Xt)^2+4L^2C_{n}^2}}^{\infty}\frac{d\tilde{Y}}{8 \pi L(e^{\tilde{Y}}-1)}\bigg]; \label{lowtemp2}
\end{split}
\ee
After interchanging the order of integration in above we get
\begin{multline}
c(\alpha)=\frac{1}{\pi\alpha}\bigg[\int_{0}^{\infty}sin(Z Y) \frac{Y^2dY}{16\pi L^3(e^{Y}-1)}+\sum_{n \in \mathbb{Z} \backslash \left\lbrace 0 \right\rbrace} \int_{2LC_{n}}^{\infty}sin\Big(Z \sqrt{\tilde{Y}^2-4L^2 C_{n}^2}\Big)\frac{\tilde{Y}^2d\tilde{Y}}{16\pi L^3(e^{\tilde{Y}}-1)} \\ - \sum_{n \in \mathbb{Z} \backslash \left\lbrace 0 \right\rbrace}C_{n}^2\int_{2LC_{n}}^{\infty}sin\Big(Z \sqrt{\tilde{Y}^2-4L^2 C_{n}^2}\Big)\frac{d\tilde{Y}}{8 \pi L(e^{\tilde{Y}}-1)}\bigg],  \label{lowtemp3}
\end{multline}
where $Z=\frac{\alpha}{t}$, $t=\frac{4 \pi L}{\beta}$. We use identity $\int_{0}^{\infty}\frac{sin(ZY)}{(e^{Y}-1)}dY= \frac{\pi}{2}coth\pi Z -\frac{1}{2Z}$ and an identity given in \cite{grad} in above equation and after straightforward simplifications we find
\begin{multline}
c(\alpha)\Big|_{\alpha=2\pi m}= -\frac{1}{32m\pi^3L^3}\bigg[4\pi^3\Big\{\frac{e^{-\frac{4\pi^2 m}{t}}(1+e^{-\frac{4\pi^2 m}{t}})}{(1-e^{-\frac{4\pi^2 m }{t}})^3}\Big\}-\frac{t^3}{(2\pi m)^3}\bigg] \\ +\frac{1}{32m\pi^3 L^3}\sum_{n \in \mathbb{Z} \backslash \left\lbrace 0 \right\rbrace}\bigg\{-\frac{d^2}{dZ^2}\bigg(\sum_{\tilde{n}=1}^{\infty}\frac{2LC_{n}Z}{\sqrt{\tilde{n}^2+Z^2}}K_{1}\Big(2LC_{n}\sqrt{\tilde{n}^2+Z^2}\Big)\bigg)\\ +\frac{1}{16m\pi^3 L}\sum_{n \in \mathbb{Z} \backslash \left\lbrace 0 \right\rbrace}C_{n}^2\bigg(\sum_{\tilde{n}=1}^{\infty}\frac{2LC_{n}Z}{\sqrt{\tilde{n}^2+Z^2}}K_{1}\Big(2LC_{n}\sqrt{\tilde{n}^2+Z^2}\Big)\bigg)\bigg]\bigg |_{Z=\frac{\alpha}{t}=\frac{2\pi m}{t}} \label{lowtemp4}
\end{multline}
Now using Eqn.(\ref{tempcas2}), Eqn.(\ref{poisson2}) and Eqn.(\ref{lowtemp4}), we get the low temperature limit of the finite thermal correction for Casimir force as
\begin{multline}
F_{\beta>>4\pi L}=-\bigg\{\frac{\pi^2}{480L^4}\Big(1+\frac{t^4}{48\pi^4}-\frac{60t}{\pi^2}e^{-\frac{4\pi^2}{t}}\Big)\bigg\}\\-\frac{1}{32\pi^2 L^2}\sum_{n \in \mathbb{Z} \backslash \left\lbrace 0 \right\rbrace}\Big\{(\sqrt{\pi L C_{n}})\Big(2L C_{n} +\frac{9}{2}\Big)e^{-2L C_{n}}\Big\}-\frac{1}{16\pi^2 L^2}\sum_{n \in \mathbb{Z} \backslash \left\lbrace 0 \right\rbrace}C_{n}^2(\sqrt{\pi L C_{n}})e^{-2L C_{n}} \\-\frac{1}{16 \pi^2 L^4}\sum_{n \in \mathbb{Z} \backslash \left\lbrace 0 \right\rbrace}(\sqrt{\pi L C_{n}})\bigg\{\frac{2LC_{n} t^{\frac{5}{2}}}{(t^2+4\pi^2)^{\frac{5}{4}}}+\frac{9}{2}\frac{ t^{\frac{7}{2}}}{(t^2+4\pi^2)^{\frac{7}{4}}}+\frac{16\pi^2L^2 C_{n}^2 t^{\frac{3}{2}}}{(t^2+4\pi^2)^{\frac{7}{4}}}-\frac{8\pi^2L C_{n} t^{\frac{5}{2}}}{(t^2+4\pi^2)^{\frac{9}{4}}}\\+\frac{21 \pi^2 t^{\frac{9}{2}}}{(t^2+4\pi^2)^{\frac{11}{4}}}\bigg\}e^{- 2LC_{n}t\sqrt{t^2+4\pi^2}}-\frac{1}{8\pi^2 L^2}\sum_{n \in \mathbb{Z} \backslash \left\lbrace 0 \right\rbrace}C_{n}^2(\sqrt{\pi LC_{n}})\bigg(\frac{ t^{\frac{3}{2}}}{(t^2+4\pi^2)^{\frac{3}{4}}}\bigg)e^{- 2LC_{n}t\sqrt{t^2+4\pi^2}}.\label{lowtemp5}
\end{multline}
Note here that at $T=0$, above force expression reduce to commutative results \cite{kam,Brevik} with additional temperature independent correction terms due to non-commutativity of space-time.
 
Note that, the commutative limit is obtained by letting $R \rightarrow 0$, where $R$ is the size of the extra compactified dimension. We find that the thermal corrections to Casimir force (both at the high and low-temperature limit) reduce to known commutative result when $R \rightarrow 0$\cite{kam,Brevik}. At the high-temperature limit, Casimir force has, apart from the $L^{-3}$ dependent corrections, $L^{-1}$ dependent corrections also. 
 Similarly, at a low-temperature limit, Casimir force has a dependency on $L^{-4}$ terms and also additional $L^{-2}$ dependent terms due to non-commutativity. Through expansion of exponential part of the correction terms, we find that in both high and low-temperature limits, Casimir force also has a $L$-independent corrections due to non-commutativity.

\section{Conclusion}

In this paper, we have studied the effect of vacuum fluctuations in $4+1$-dimensional DFR space-time by analyzing the Casimir effect between parallel plates. We have found modifications to the Casimir force and the Casimir energy between two parallel plates due to the non-commutativity of DFR space-time. Here we found that if the additional $\theta$-directions are treated on the same footing as transverse commutative directions, the Casimir force (and energy) found to be exactly same as that for commutative extra dimensional space-time \cite{kam} and does not have dependence on the non-commutative parameters. We then evaluated the Casimir effect by treating the non-commutative direction as compactified dimension. We find that the Casimir force and energy becomes complex if the weight function $W(\theta)$ introduced in Eqn.(\ref{spproduct}) is not a constant. Hence we take $W(\theta)=1$ as in the studied in \cite{amo5} in our analysis. In this case, we have seen that Casimir force proportional to $\frac{1}{L^4}$ with modifications that vary as $\frac{1}{L^2}$ and $\frac{1}{L}$ in $4+1$ dimensional DFR space-time. And Casimir energy vary as $\frac{1}{L^3}$ with correction term due to non-commutativity of space-time vary as $\frac{1}{L}$. This result is in agreement with similar results obtained in \cite{prd79,plb668,farina} for Casimir effect in commutative space-time with extra compactified-dimension. But our result is in contrast with the results of \cite{kam}, where Casimir force and Casimir energy scale as $\frac{1}{L^5}$ and $\frac{1}{L^4}$, respectively in $4+1$ dimensions, where extra dimension was a commutative one. 

In \cite{prd79,plb668,farina}, the Casimir effect was studied in the presence of extra dimension and found that Casimir force has terms that vary as $\frac{1}{L^4}$ apart from correction terms that vary as $\frac{1}{L^2}$ and $\frac{1}{L}$ due to presence of one extra dimension in addition to $4$ dimensional commutative Minkowski space-time. In \cite{kam}, Casimir effect is studied for parallel conducting plates in arbitrary spatial dimension D and showed that Casimir force and Casimir energy scale as $L^{-(D+1)}$ and $L^{-(D)}$ respectively. For spatial dimension $D=4$, these results are in contrast with the results obtained in \cite{prd79,plb668,farina}. This difference in the L dependence observed in \cite{prd79,plb668,farina} and \cite{kam} is because of the fact that in \cite{kam}, the extra dimension is treated as transverse direction to the plates, whereas in \cite{prd79,plb668,farina}, the effect of the extra dimension on parallel plates in usual $3+1$ dimensional Minkowski space-time is studied. The approach taken in \cite{prd79,plb668,farina} is similar to our present study. 

Note that we have obtained the corrections to Casimir force that is attractive in nature (see Eqn.(\ref{forcefinal2})). We have showed that in commutative limit, $R \rightarrow 0$  our results reduce to that in \cite{Mil}. Here $R$ is the size of the extra compactified dimension - $\theta$. Note that the Casimir force expression in Eqn.(\ref{forcefinal2}) reduces to $-\frac{\pi^2}{480L^4}$ in the commutative limit. Casimir force expression (see Eqn.(\ref{forcefinal2}))has correction terms that vary as $\frac{1}{R^{2}}$ and $\frac{1}{R^3}$. In this Casimir force expression, except correction term with $n=1$ and $m=1$, contribution of correction terms with higher values of $n,m$, are very small. Thus to exhibit the generic nature of the Casimir force, we neglect those terms and for various possible values of size of the extra compactified dimension, $R$, we see how the Casimir force is varying with plate separation (see the plot).

\begin{figure}[H]
\centering
\includegraphics[scale=.70]{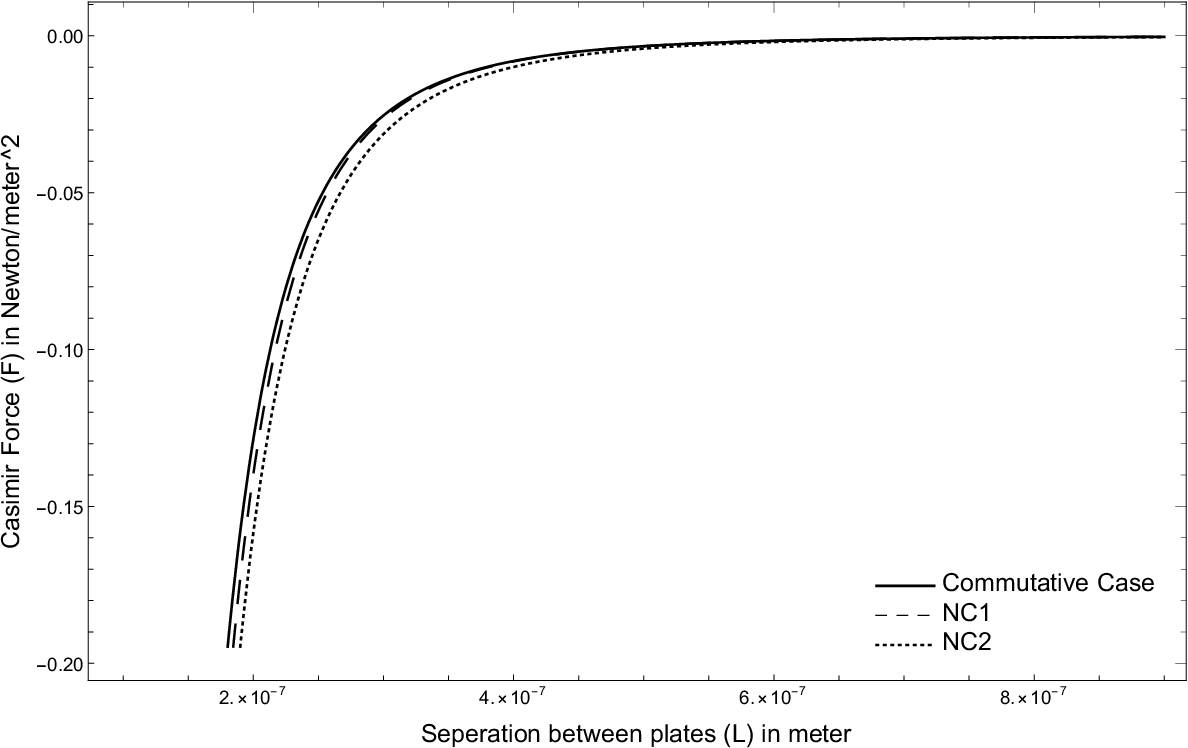} 
\caption{Variation of Casimir force with plate separation.}
\end{figure}

In the above plot, NC1 is the plot for $ R \sim 10^{-7} m$ and NC2 is for $ R \sim 10^{-5}m $. The solid line gives the Casimir force for commutative case.

Here the solid, continuous line shows the variation of the Casimir force in the commutative $4$-dimensional space-time for parallel plates as a function of plate separation. For $R<10^{-7}m$, corrections in the Casimir force, are very small. All the plots overlap with solid line (Casimir force for the commutative case) for large plate separation. The Casimir force in DFR space-time, for values of $R \sim 10^{-7}m$  and $R \sim 10^{-5}m$ are larger than the commutative case for a given plate seperation. That is, modifications to the Casimir force strengthen the attractive nature of the Casimir force. But for $R<10^{-7}m$, as corrections to the Casimir force are very small that we do not find any deviation from the commutative case in the plot. Thus the present analysis constrain the size of the extra compactified dimension with lower bound, i.e., $R \geq 10^{-7}m$. 

We have also investigated finite temperature corrections to the Casimir effect, in presence of extra compactified dimension, in $4+1$ dimensional DFR space-time. For the high-temperature limit, we have shown that the Casimir force has terms that scale as $\frac{1}{L^3}$ and $\frac{1}{L}$ respectively. $\frac{1}{L}$ term is purely due to the non-commutativity of space-time. Similarly, we have studied the low-temperature limit of the Casimir force, and it is shown to have terms that vary as $\frac{1}{L^4}$ and $\frac{1}{L^2}$. Here $\frac{1}{L^2}$ is purely due to the non-commutativity of space-time. In \cite{rype,teo1}, finite temperature correction to Casimir effect due to the presence of one extra commutative dimension was studied and the result shows the same feature we find here in DFR space-time. 

In the appendix, we have investigated the Casimir effect for $3+1$ dimensional massive scalar field theory \cite{kam} obtained from $7$-dimensional massless DFRA scalar field theory in DFR space-time by applying the Kaluza-Klein reduction method.

\section*{Acknowledgements}
SKP thanks UGC, India, for support through the JRF scheme(id.191620059604).

\bigskip

\renewcommand{\thesection}{Appendix : A}
\section{Kaluza-Klein reduction and Casimir effect}
\renewcommand{\thesection}{A}
In this appendix, we study the Casimir effect for massive scalar field theory in 3+1 dimensional Minkowski space-time. This model is obtained by applying the Kaluza-Klein reduction prescription to 7-dimensional DFRA massless scalar field theory.

We start with the action of massless complex scalar field theory in DFR space-time,
\be
S=\int d^4x~d^3\theta~W(\theta)\bigg(\partial_{\mu}\phi(x,\theta)\partial^{\mu}\phi^{\dagger}(x,\theta)+\partial_{\theta^i}\phi(x,\theta)\partial^{\theta^i}\phi^{\dagger}(x,\theta)\bigg). \label{complex-action}
\ee
We rewrite this space-time as direct product space, $M_{7}=M_{4}\otimes K_{3}$, where $K_{3}$ is a compact space. Using Kaluza-Klein reduction, from this 7-dimensional massless scalar field theory, we obtain a massive scalar field theory in $3+1$ dimensional Minkowski space-time. For this we take Fourier decomposition of DFRA complex scalar field as,

\be 
\phi(x,\theta)=\sum_{N=-\infty}^{\infty}\phi^{(N)}(x)\exp{\Big(-\frac{iNM\theta}{\lambda}\Big)}. \label{kal-fourier1}
\ee
Using the above in Eq.(\ref{complex-action}), we get 
\be 
S=\sum_{N=-\infty}^{\infty}\int d^4x~d^3\theta~W(\theta)\bigg(\partial_{\mu}\phi^{(N)}(x)\partial^{\mu}\phi^{\dagger~(N)}(x)+(N^2M^2)\phi^{(N)}(x)\phi^{\dagger~(N)}(x)\bigg). \label{kal-fourier2}
\ee
Now after integrating over $\theta$ from $0$ to $2\pi/\lambda$ , we find
\be 
S=\sum_{N=-\infty}^{\infty}\int d^4x~\Big(erf\Big(\frac{\pi}{\lambda^3}\Big)\Big)^3\bigg(\partial_{\mu}\phi^{(N)}(x)\partial^{\mu}\phi^{\dagger~(N)}(x)+(N^2M^2)\phi^{(N)}(x)\phi^{\dagger~(N)}(x)\bigg), \label{kal-red}
\ee
where $erf()$ is the error function.
Next for $\lambda\rightarrow 0$, $erf(\frac{\pi}{\lambda^3})\rightarrow 1$ the above equation reduces to
\be 
S=\sum_{N=-\infty}^{\infty}\int d^4x~\bigg(\partial_{\mu}\phi^{(N)}(x)\partial^{\mu}\phi^{\dagger~(N)}(x)+(N^2M^2)\phi^{(N)}(x)\phi^{\dagger~(N)}(x)\bigg). \label{kaluza-action}
\ee
Thus we get the action of an infinite number of uncoupled modes of the massive complex scalar field in the usual Minkowski space-time. Using this, we can study the Casimir effect, and for single-mode, the corresponding Casimir force and Casimir energy expressions \cite{kam} are
\be
F_{c}=-\frac{1}{32\pi^2 L^4}\bigg[\frac{3}{4}\sum_{n=1}^{\infty}\frac{(4ML)^2}{n^2}K_{2}(2nML)+2M^2L^2\sum_{n=1}^{\infty}\Big(\frac{4ML}{n}\Big)K_{1}(2nML)\bigg] 
\ee
and 
\be
E_{c}=-\frac{1}{32\pi^2 L^3}\sum_{n=1}^{\infty}\frac{(2ML)^2}{n^2}K_{2}(2nML) 
\ee
respectively where $K_{1}$ and $K_{2}$ are modified Bessel functions.
These results match exactly with the results of the Casimir force for massive scalar field theory given in \cite{kam}.
Thus we show that one can obtain the usual massive scalar field theory by applying Kaluza-Klein reduction to the massless DFRA scalar field theory, and this will lead to the Casimir effect as in the well studied massive scalar theory in $3+1$ dimensions.


\end{document}